\documentclass[twoside,twocolumn]{article}
\usepackage[utf8]{inputenc}
\usepackage{booktabs}
\usepackage[table,x11names]{xcolor}
\usepackage{graphicx}
\usepackage{pdfpages}
\usepackage{blindtext}
\usepackage{float} 
\usepackage[url=false,uniquename=false,uniquelist=false,date=year,natbib=true,doi=false,style=nature]{biblatex}
\usepackage[hidelinks]{hyperref}
\usepackage[scaled=1.2]{fbb}
\usepackage[scaled=1.2]{inconsolata}
\usepackage[libertine,scaled=1.2]{newtxmath}
\usepackage[cal=boondoxo,bb=boondox,frak=boondox]{mathalpha}
\usepackage{titling}
\usepackage{abstract}
\usepackage[]{enumitem} 
\usepackage{setspace} 

\usepackage[margin=1.2in]{geometry}
\usepackage{fancyhdr}
\usepackage[skip=2pt plus1pt]{parskip}
\usepackage{titlesec}
\titlespacing\section{0pt}{12pt plus 4pt minus 2pt}{0pt plus 2pt minus 2pt}

\newcommand*\samethanks[1][\value{footnote}]{\footnotemark[#1]}

\makeatletter
\newcommand*{\centerfloat}{%
  \parindent \z@
  \leftskip \z@ \@plus 1fil \@minus \textwidth
  \rightskip\leftskip
  \parfillskip \z@skip}
\makeatother

\title{Understanding AI alignment research: A Systematic Analysis}
\author{Jan H.~Kirchner\thanks{These authors contributed equally.} \\
  \small{\texttt{kirchner.jan@icloud.com}}  \and Logan Smith\samethanks[1]{} \\
  \small{\texttt{logansmith5@gmail.com}} \and Jacques Thibodeau\samethanks[1]{} \\
  \small{\texttt{thibo.jacques@gmail.com}} \and Kyle McDonell \\
  \small{\texttt{kyle@conjecture.dev}} \and Laria Reynolds\\
  \small{\texttt{laria@conjecture.dev}}
  }%
\date{}

\begin{document}
\twocolumn[
  \begin{@twocolumnfalse}
    \maketitle
    \vspace{-0.5cm}
    \begin{abstract} \small
        \vspace{0.25cm}
    AI alignment research is the field of study dedicated to ensuring that artificial intelligence (AI) benefits humans. As machine intelligence gets more advanced, this research is becoming increasingly important. Researchers in the field share ideas across different media to speed up the exchange of information. However, this focus on speed means that the research landscape is opaque, making it difficult for young researchers to enter the field. In this project, we collected and analyzed existing AI alignment research. We found that the field is growing quickly, with several subfields emerging in parallel. We looked at the subfields and identified the prominent researchers, recurring topics, and different modes of communication in each. Furthermore, we found that a classifier trained on AI alignment research articles can detect relevant articles that we did not originally include in the dataset. We are sharing the dataset with the research community and hope to develop tools in the future that will help both established researchers and young researchers get more involved in the field.\\
\end{abstract}
  \end{@twocolumnfalse}
]
\saythanks
\section*{Introduction}
\textit{AI alignment research} is a nascent field of research concerned with developing machine intelligence in ways that achieve desirable outcomes and avoid adverse outcomes\autocite{yudkowsky2016ai,christian2020alignment}. While the term \textit{alignment problem} was originally proposed to denote the problem of "pointing an AI in a direction"\autocite{yudkowsky_rocket_2018}, the term \textit{AI alignment research} is now used as an overarching term referring to the entire research field associated with this problem\autocite{christian2020alignment,russell2021human,gabriel2020artificial,ouyang2022training,kenton2021alignment,dafoe2021cooperative,askell2021general}. Associated lines of research include the question of how to infer human values as revealed by preferences\autocite{christiano2017deep}, how to prevent risks from learned optimization\autocite{hubinger2019risks}, or how to set up an appropriate structure of governance to facilitate coordination\autocite{dafoe2018ai}.

As machine intelligence becomes increasingly capable\autocite{grace2018will,sevilla2022compute}, AI alignment research becomes increasingly important. There is a risk that if machine intelligence is not carefully designed, it could have catastrophic consequences for humanity\autocite{bostrom2017superintelligence,ord2020precipice,carlsmithpower}. For example, if machine intelligence is not designed to take human values into account, it could make decisions that are harmful to humans\autocite{bostrom2017superintelligence}. Alternatively, if machine intelligence is not designed to be transparent and understandable to humans, it could make decisions that are opaque to humans and difficult to understand or reverse\autocite{christiano2019What}. As machine intelligence rapidly becomes more powerful\autocite{sevilla2022compute}, the stakes associated with the AI alignment problem only grow. Consequently, the field receives considerable attention from philanthropic organizations searching to increase the speed and scope of research\autocite{OpenPhil, FutureFund}.

One interesting feature of AI alignment research is how the researchers communicate: to increase the speed and bandwidth of information exchange, novel insights and ideas are exchanged across various media. Beyond the traditional research article published as a preprint or conference article, a substantial portion of AI alignment research is communicated on a curated community forum: the Alignment Forum\autocite{AlignmentForum}. Other channels of communication include formal and informal talks\autocite{Christiano2020Current}, semi-publicly shared manuscripts and notes\autocite{Cotra2020Draft,carlsmithpower}, and informal exchanges via instant messaging\autocite{MIRIConversations}.

The strong focus on increased speed and bandwidth of communication comes at the cost of a diffuse research landscape, making it difficult for newcomers to orient themselves\autocite{Hyvaerinen2022How,wentworth2021How}. These difficulties are exacerbated by the short time the field has existed and the resulting lack of unifying paradigms\autocite{kuhn1970structure,shimi2021Epistemological}. Previous attempts to catalog and classify existing AI alignment research\autocite{StampyWiki, AlignmentNewsletter, AISafetyPapers, ElicitAssistant} do not include all relevant sources, are not kept up-to-date, and do not provide easy access to the data in a machine-readable format. Given the potential importance of AI alignment research and the attempts to increase the size of the field\autocite{OpenPhil, FutureFund}, the lack of a coherent overview of the research landscape represents a major bottleneck.

In this project, we collected and cataloged AI alignment research literature and analyzed the resulting dataset in an unbiased way to identify major research directions. We found that the field is growing rapidly, with several subfields emerging naturally over time. By analyzing the emerging subfields, we can identify the prominent researchers working in the subfield, recurring topics and questions specific to each subfield, and different modes of communication dominating each subfield. Finally, training a classifier to distinguish AI alignment research from more general AI research can automatically detect relevant articles published too recently to be included in our dataset. We make our dataset and the analysis publicly available to interested researchers to enable further analysis and facilitate orientation to the field.

\section*{Results}

To capture the current state of AI alignment research, we collected research articles from various sources (Tab.~\ref{tab:dataset}). Beyond the full-length manuscript published on arXiv ($N=707$), we also included shorter communications published on the Alignment Forum ($N=2,138$), blogs, and personal websites ($N=1,326$),  publicly available, full-length books ($N=23$), a popular AI alignment research newsletter with summaries of articles ($N=420$), full-length manuscripts not published on arXiv ($N=372$), transcripts of lectures and interviews ($N=494$), and entries from public wikis ($N=582$).  To establish a baseline for our analysis, we also collected research articles from adjacent ($N=1,679$) and unrelated ($N=1,000$) areas of research, as well as shorter communications published on the LessWrong Forum ($N=28,259$). For details about our collection procedure, see the Methods section.

{\renewcommand{\arraystretch}{1.4}
\setlength\arrayrulewidth{0.5pt}
\begin{table*}[]
    \small
    \centerfloat
    \begin{tabular}{p{4cm}|p{9cm}|c}
    \toprule        
    \textbf{source} & \textbf{domain} & \textbf{\# of articles} \\  \midrule \midrule
     \rowcolor{green!10}  \textbf{Alignment Forum}  & \href{https://www.alignmentforum.org/}{\texttt{alignmentforum.org}} & \texttt{2,138} \\ 
      \rowcolor{lightgray!20} & \href{https://www.lesswrong.com/}{\texttt{lesswrong.com}} & \texttt{28,252} \\ \hline
       \rowcolor{green!10} \textbf{arXiv} & \texttt{AI alignment research (level-0)} & \texttt{707} \\ 
       \rowcolor{lightgray!20} & \texttt{AI research (level-1)}  & \texttt{1,679} \\ 
        \rowcolor{lightgray!20} & \href{https://arXiv.org/search/?query=quantum&searchtype=all&source=header}{\texttt{arXiv.org/search/?query=quantum}} & \texttt{1,000}   \\
        \rowcolor{purple!10} & \href{https://arXiv.org/list/cs.AI/recent}{\texttt{arXiv.org/list/cs.AI }}\texttt{(filtered)}  & \texttt{4,621}   \\ \hline
       \textbf{Books}  & \texttt{(available upon request)} & \texttt{23} \\ \hline
       \textbf{Blogs}  & \href{https://aiimpacts.org/}{\texttt{aiimpacts.org}} & \texttt{227} \\ 
         & \href{https://aipulse.org/}{\texttt{aipulse.org}} & \texttt{23} \\ 
         & \href{https://aisafety.camp/}{\texttt{aisafety.camp}} & \texttt{8} \\ 
         & \href{https://carado.moe/}{\texttt{carado.moe}} & \texttt{59} \\ 
         & \href{https://www.cold-takes.com/}{\texttt{cold-takes.com}}& \texttt{111} \\ 
         & \href{https://deepmindsafetyresearch.medium.com/}{\texttt{deepmindsafetyresearch.medium.com}}& \texttt{10} \\
         & \href{https://generative.ink/}{\texttt{generative.ink}} & \texttt{17} \\ 
         & \href{https://www.gwern.net/}{\texttt{gwern.net}} & \texttt{7} \\ 
         & \href{https://intelligence.org/}{\texttt{intelligence.org}} & \texttt{479} \\ 
         & \href{https://jsteinhardt.wordpress.com/}{\texttt{jsteinhardt.wordpress.com}}& \texttt{39} \\ 
         & \href{https://qualiacomputing.com/}{\texttt{qualiacomputing.com}}& \texttt{278} \\ 
         & \href{https://vkrakovna.wordpress.com/}{\texttt{vkrakovna.wordpress.com}}& \texttt{43} \\
         & \href{https://waitbutwhy.com/2015/01/artificial-intelligence-revolution-1.html}{\texttt{waitbutwhy.com}}& \texttt{2} \\
         & \href{https://www.yudkowsky.net/}{\texttt{yudkowsky.net}} & \texttt{23} \\ \hline
       \textbf{Newsletter}  & \href{https://rohinshah.com/alignment-newsletter/}{\texttt{rohinshah.com/alignment-newsletter/ }} \texttt{summaries} & \texttt{420} \\ \hline
       \textbf{Reports}  & \texttt{pdf-only articles} & \texttt{323} \\
        & \href{https://distill.pub/}{\texttt{distill.pub}}& \texttt{49} \\ \hline
       \textbf{Audio transcripts}  & \href{https://www.youtube.com/playlist?list=PLTYHZYmxohXp0xvVJmMmpT_eFJovlzn0l}{\texttt{youtube.com playlist 1}} \& \href{https://www.youtube.com/playlist?list=PLTYHZYmxohXpn5uf8JZ2OouB1PsDJAk-x }{\texttt{2}}& \texttt{457} \\ 
       & \texttt{Assorted transcripts} & \texttt{25} \\
       & \texttt{interviews with AI researchers}\autocite{Gates2022Transcripts} & \texttt{12} \\  \hline
       \textbf{Wikis} & \href{https://arbital.com/}{\texttt{arbital.com}}& \texttt{223} \\
       & \href{https://www.lesswrong.com/tags/all}{\texttt{lesswrong.com (Concepts Portal)}} & \texttt{227} \\
       & \href{https://stampy.ai/wiki/Stampy\%27s_Wiki}{\texttt{stampy.ai}} & \texttt{132} \\ \hline\hline
       \textbf{Total:} & \texttt{Total token count: 89,240,129} &   \\
       & \texttt{Total word count: 53,550,146} &   \\
       & \texttt{Total character count: 351,757,163} &   \\\bottomrule
    \end{tabular}
    \caption{\textbf{Different sources of text included in the dataset alongside the number of articles per source.} Color of row indicates that data was analyzed as AI alignment research articles (green) or baseline (gray), or that the articles were added to the dataset as a result of the analysis in Fig.~\ref{fig:filter} (purple). Definition of level-0 and level-1 articles in Fig.~\ref{fig:filter}c. For details about our collection procedure see the Methods section.}
    \label{tab:dataset}
\end{table*}}

\noindent
\subsection*{Rapid growth of AI alignment research from 2012 to 2022 across two platforms.}

There was substantial heterogeneity in the form and quality of articles in the dataset. We decided to focus on articles published on the Alignment Forum and as preprints on the arXiv server (see Methods for arXiv inclusion criteria). These sources contain a large portion of the entire published AI alignment research (Tab.~\ref{tab:dataset}) and are structured in a consistent form that allows automated analysis.

\begin{figure*}
    \centering
    \includegraphics{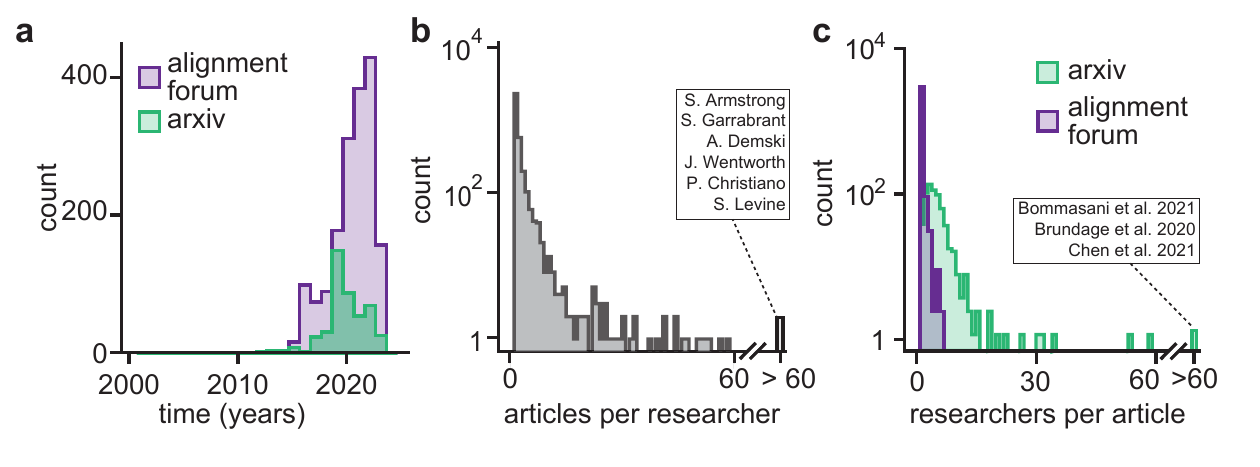}
    \caption{\textbf{Alignment research across a community forum and a preprint server.} (\textbf{a}) Number of articles published as a function of time on the Alignment Forum (AF; purple) and the arXiv preprint server (arXiv; green). (\textbf{b}) Histogram of the number of articles per researcher published on either AF or arXiv. Inset shows names of six researchers with more than 60 articles. Note the logarithmic y-axis. (\textbf{c}) Histogram of the number of researchers per article on AF (purple) and arXiv (green). Note the logarithmic y-axis. }
    \label{fig:descriptive}
\end{figure*}

To quantify the field's growth, we visualized the number of articles published on either platform as a function of time. We found a rapid increase from 2017\footnote[3]{We note that the Alignment Forum was created in 2018\autocite{Raemon2018Announcing}.} to 2022 (present) from less than 20 articles per year to over 400 (Fig.~\ref{fig:descriptive}a). When calculating the number of articles published per researcher, we observed a long-tailed distribution with most researchers publishing less than five articles and some publishing more than 60 (Fig.~\ref{fig:descriptive}b). Finally, when comparing the number of researchers per article on the Alignment Forum and the arXiv, we noticed that articles on the Alignment Forum tend to be written by either just a single author or by a small team of fewer than five researchers (Fig.~\ref{fig:descriptive}c; purple). In contrast, the distribution of authors on arXiv articles is long-tailed and includes articles with more than 60 authors\autocite{bommasani2021opportunities, brundage2020toward, chen2021evaluating} (Fig.~\ref{fig:descriptive}c; green). This asymmetry partially results from the late introduction of the multiple authors feature to the Alignment Forum\footnote{The feature to add multiple authors didn't become available to all users until 2019, and many people may still not be aware of how to do it. }, but might also reflect the Alignment Forum's focus on speed of communication, which disincentivizes large collaborations\autocite{moshontz2021guide}. Alternatively, the larger number of authors on arXiv articles might also reflect inflation of (unjustified) authorship on research articles\autocite{poder2010let, lozano2013elephant}.

Thus, AI alignment research is a rapidly growing field, driven by many researchers contributing individual articles and a few publishing prolifically.

\subsection*{Unsupervised decomposition of AI alignment research into distinct clusters.}

\begin{figure*}
    \centerfloat
    \includegraphics{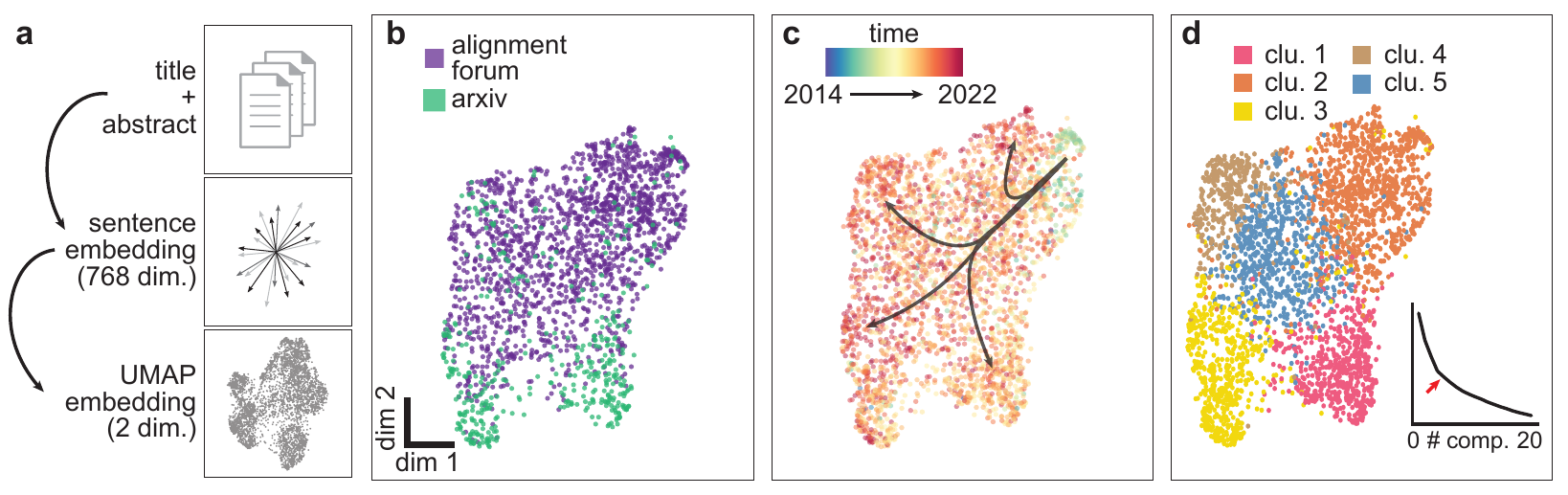}
    \caption{\textbf{Dimensionality reduction and unsupervised clustering of alignment research.} (\textbf{a}) Schematic of the embedding and dimensionality reduction. After concatenating title and abstract of articles, we embed the resulting string with the Allen SPECTER model \autocite{cohan2020specter}, and then perform UMAP dimensionality reduction with \texttt{n\_neighbors=250}. (\textbf{b}) UMAP embedding of articles with color indicating the source (AF, purple; arXiv, green). (\textbf{c}) UMAP embedding of articles with color indicating date of publication. Arrows superimposed to indicate direction of temporal evolution. (\textbf{d}) UMAP embedding of articles with color indicating cluster membership as determined with k-means (\texttt{k=5}). Inset shows sum of residuals as a function of clusters \texttt{k}, with an arrow highlighting the chosen number of clusters.}
    \label{fig:clustering}
\end{figure*}

Given the collected AI alignment research articles from the Alignment Forum and arXiv, we were curious whether we could use the text to understand the current state of research. To this end, we used the Allen SPECTER model\autocite{cohan2020specter} to compute a sentence embedding, followed by a UMAP projection\autocite{mcinnes2018umap} to obtain a low-dimensional representation (Fig.~\ref{fig:clustering}a). While there is a tendency for articles from different sources to occupy different regions of the embedding, the transition between Alignment Forum and arXiv is fluent (Fig.~\ref{fig:clustering}b). Interestingly, when visualizing the publication date, we noticed that the embedding captures part of the temporal evolution of the field (Fig.~\ref{fig:clustering}c).

Due to the relative youth of the field, there is no universally-accepted decomposition of AI alignment research into subfields\autocite{shimi2021Epistemological, Christiano2020Current, Critch2020Some}. To see if we can produce a useful, unbiased decomposition of the research landscape, we applied k-means clustering to the SPECTER embedding to obtain five distinct clusters (see Methods for details).

In summary, combined semantic embedding and dimensionality reduction produce a compact visualization of AI alignment research.

\subsection*{Research dynamics vary across the identified clusters.}

\renewcommand{\arraystretch}{1.75}
\setlength\arrayrulewidth{0.5pt}

\begin{table*}[]
\centerfloat
\small
\begin{tabular}{@{}p{3.3cm}|p{3.3cm}|p{3.3cm}|p{3.3cm}|p{3.3cm}@{}} \toprule
\textbf{cluster 1}; $N = 567$ & \textbf{cluster 2}; $N = 988$ & \textbf{cluster 3}; $N = 593$  & \textbf{cluster 4}; $N = 383$  & \textbf{cluster 5}; $N = 670$   \\[-1.5ex] (\textit{agent alignment})  & (\textit{alignment foundations}) & (\textit{tool alignment}) & (\textit{AI governance}) &  (\textit{value alignment}) \\ \midrule \midrule
S. Levine (55)           & S. Armstrong (154)     & J. Steinhardt (20)    & D. Kokotajlo (21)       & S. Armstrong  (54) \\ \hline
P. Abbeel (34)           & S. Garrabrant (95)     & D. Hendrycks (17)     & A. Dafoe (19)            & S. Byrnes (32)  \\ \hline
A. Dragan (29)           & A. Demski (94)         & E. Hubinger (14)      & G. Worley III (11)      & P. Christiano (29)      \\ \hline
S. Russell (23)          & J. Wentworth (57)      & P. Christiano (13)    & J. Clarck (10)          & R. Ngo (25)   \\ \hline
S. Armstrong (22)        & "Diffractor" (44)      & P. Kohli (11)         & S. Armstrong (9)        & R. Shah (25)    \\\bottomrule
 \hline
\end{tabular}
\caption{\textbf{Researchers with the highest number of articles per cluster.} Clusters as determined in Fig.~\ref{fig:clustering}, with number of articles per cluster $N$. Number in brackets behind researcher name indicates number of articles published by that researcher. Note: "Diffractor" is an undisclosed pseudonym.} \label{tab:researcher}
\end{table*}

\begin{figure*}
    \centerfloat
    \includegraphics{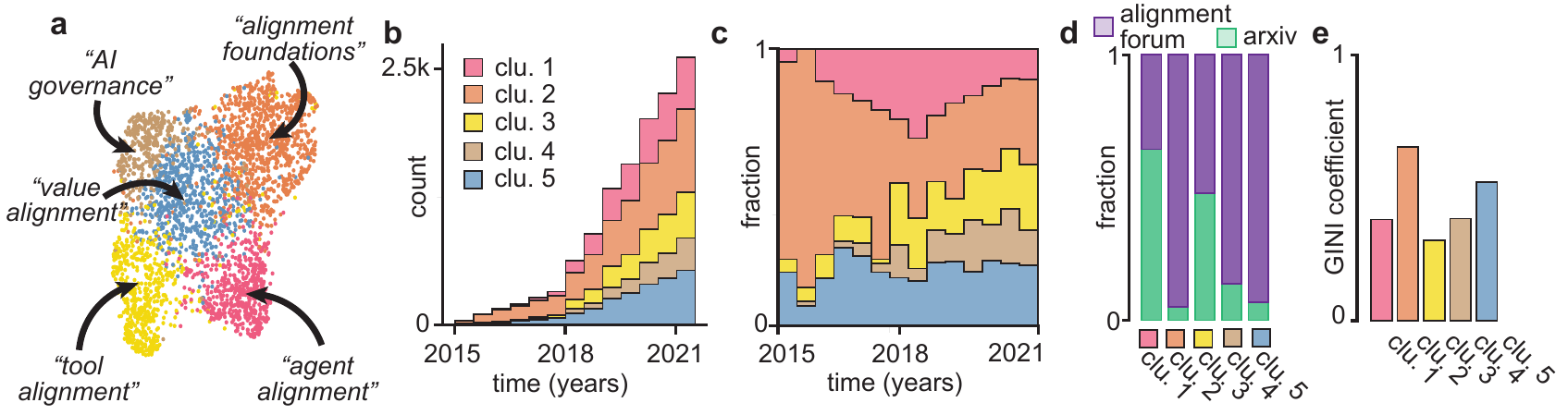}
    \caption{\textbf{Characteristics of research clusters corroborate potential usefulness of decomposition.} (\textbf{a}) UMAP embedding of articles with color indicating cluster membership as in Fig.~\ref{fig:clustering}d. Labels assigned to each cluster are putative descriptions of a common research focus across articles in the cluster. (\textbf{b}) Number of articles published per year, colored by cluster membership. (\textbf{c}) Fraction of articles published by cluster membership as a function of time. (\textbf{d}) Fraction of articles from AF or arXiv as a function of cluster membership. (\textbf{e}) GINI inequality coefficient of articles per researcher as a function of article cluster membership.}
    \label{fig:interpret}
\end{figure*}

Having identified five distinct research clusters, we asked ourselves if we could find natural descriptions of research topics and prominent researchers. Therefore, we inspected which researchers tend to publish the highest number of articles in each cluster (Tab.~\ref{tab:researcher}). Even though the names of researchers did not enter into the Allen SPECTER sentence embedding (Fig.~\ref{fig:clustering}a), we observed that different researchers tend to dominate different research clusters. The distribution of researchers across clusters lead us to assign putative labels to the clusters (Fig.~\ref{fig:interpret}a):

\begin{itemize}[leftmargin=0.2cm]
    \item \textbf{cluster one} : \textit{Agent alignment} is concerned with the problem of aligning agentic systems, i.e.~those where an AI performs actions in an environment and is typically trained via reinforcement learning.
    \item \textbf{cluster two} : \textit{Alignment foundations} is concerned with \textit{deconfusion} research, i.e.~the task of establishing formal and robust conceptual foundations for current and future AI alignment research.
    \item \textbf{cluster three} : \textit{Tool alignment} is concerned with the problem of aligning non-agentic (tool) systems, i.e.~those where an AI transforms a given input into an output. The current, prototypical example of tool AIs is the "large language model"\autocite{weidinger2021ethical,bommasani2021opportunities}.
    \item \textbf{cluster four} : \textit{AI governance} is concerned with how humanity can best navigate the transition to advanced AI systems. This includes focusing on the political, economic, military, governance, and ethical dimensions\autocite{dafoe2018ai}.
    \item \textbf{cluster five} : \textit{Value alignment} is concerned with understanding and extracting human preferences and designing methods that stop AI systems from acting against these preferences.
\end{itemize}

To corroborate these putative labels, we computed a word cloud representation of the articles (Sup.~Fig.~\ref{fig:wordcloud}). We found the recurring words specific to each cluster to be in good agreement with the labels. We also note that our labels are consistent with our observation that alignment foundations research is the historical origin of AI alignment research (Fig.~\ref{fig:clustering}c, Fig.~\ref{fig:interpret}b,c). Furthermore, we observe that theoretical research (alignment foundations, value alignment, AI governance) tends to be published on the Alignment Forum. In contrast, applied research (agent alignment, tool alignment) tends to be published on arXiv (Fig.~\ref{fig:clustering}b, Fig.~\ref{fig:interpret}d). Finally, we note that in the alignment foundations cluster, a few individual researchers tend to produce a disproportionate number of research articles (Fig.~\ref{fig:interpret}e).

In combination, these arguments make us hopeful that our unsupervised decomposition of AI alignment research mirrors relevant structures existing in the field. We hope to leverage the decomposition to provide researchers structured access to the existing literature in future work.

\subsection*{Leveraging dataset to train an AI alignment research classifier.}

When quantifying the number of articles across different sources, we noticed a dramatic drop-off in articles published on the arXiv after 2019 (Fig.~\ref{fig:descriptive}a). Especially in contrast with the continued strong increase in articles published on the Alignment Forum, we suspected that our data collection might have missed some more recent, relevant work\footnote[1]{In particular, for our dataset, we manually extended an existing collection of arXiv articles from 2020\autocite{AISafetyPapers}, see Methods section for details.}. 

\begin{figure*}
    \centerfloat
    \includegraphics{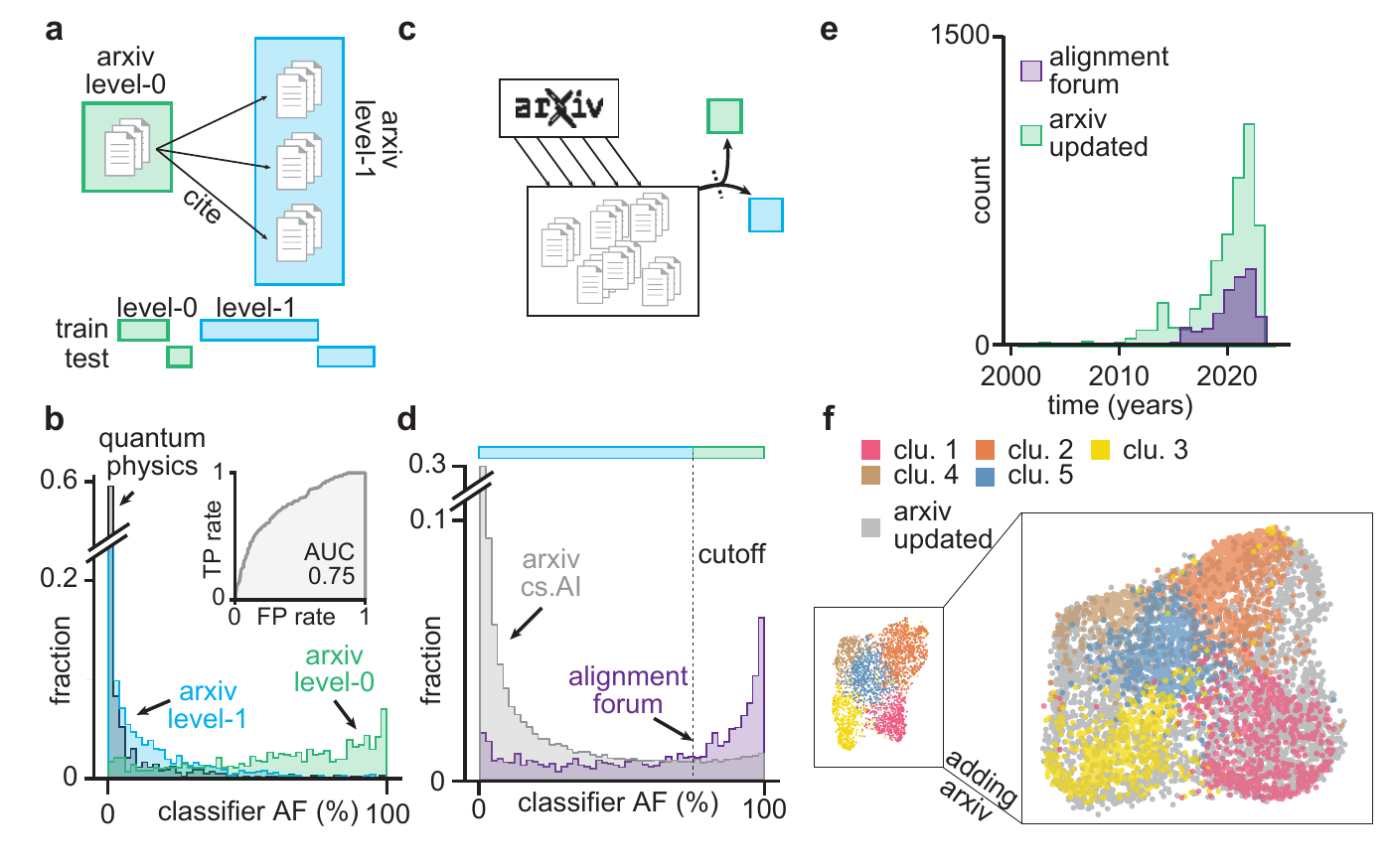}
    \caption{\textbf{An AI alignment research classifier for filtering new publications.} (\textbf{a}) Top: Illustration of arXiv level-0 articles (alignment research; green) and level-1 articles (cited by alignment research articles; blue). Bottom: Schematic of test-train split (\texttt{20\%-80\%} for training of a logistic regression classifier. (\textbf{b}) Fraction of articles as a function of classifier score for arXiv level-0 (green), level-1 (blue), and arXiv articles on quantum physics (grey). (\textbf{c}) Illustration of procedure for filtering arXiv articles. After querying articles from the \texttt{cs.AI} section of arXiv, the logistic regression classifier assigns a score between $0$ and $1$. (\textbf{d}) Fraction of articles as a function of classifier score for articles from the \texttt{cs.AI} section of arXiv (grey) and AF (purple). Dashed line indicates cutoff for classifying articles as arXiv level-0 (\texttt{75\%}). (\textbf{e}) Number of articles published as a function of time on AF (purple) and arXiv (green), according to the cutoff in panel \textbf{d}. (\textbf{f}) Left inset: Original UMAP embedding from Fig.~\ref{fig:clustering}d. Right: UMAP embedding of all original articles and updated arXiv articles with color indicating cluster membership as in Fig.~\ref{fig:clustering}d or that the article is filtered from the arXiv (gray).}
    \label{fig:filter}
\end{figure*}

To automatically detect articles published more recently, we decided to train a logistic regression classifier on the semantic embeddings of arXiv articles. Besides the AI alignment research articles already included in our dataset ("arXiv level-0"; Fig.~\ref{fig:filter}a green), we also collected all arXiv articles cited by level-0 articles, which were not level-0 articles themselves ("arXiv level-1"; Fig.~\ref{fig:filter}a blue). We trained the classifier on a training set ($80\%$) to distinguish level-0 from level-1 articles and evaluated performance on a separate test set ($20\%$). The classifier achieved good performance (AUC$=0.75$; Fig.~\ref{fig:filter}b inset), reliably rejecting level-1 articles and correctly identifying a large portion of level-0 articles (Fig.~\ref{fig:filter}b). To test whether the classifier robustly generalizes beyond AI research, we tested it on 1000 recently published articles on quantum physics and the Alignment Forum. We found that the classifier reliably rejects quantum physics and accepts Alignment Forum articles (Fig.~\ref{fig:filter}b,d).

Most AI alignment research articles on the arXiv are published in the \texttt{cs.AI} section. Therefore we used the arXiv API\autocite{arxivAPI} to collect all articles from that section (Fig.~\ref{fig:filter}c). When applying our classifier to the semantic embeddings of the \texttt{cs.AI} articles, we observed a slightly bimodal distribution with most articles receiving a score close to $0\%$, and some articles receiving a score close to $100\%$ (Fig.~\ref{fig:filter}d). Motivated by the distribution of scores of Alignment Forum articles and by individual inspection, we chose a threshold at $75\%$ and considered articles above that threshold as AI alignment research-relevant and added them to our dataset. As anticipated, we found that the number of AI alignment-relevant arXiv articles increases as rapidly over time as the articles published on the Alignment Forum (Fig.~\ref{fig:filter}e). Finally, to verify that the addition of AI alignment-relevant arXiv articles does not affect our unsupervised decomposition, we repeated the UMAP dimensionality reduction on the updated dataset. We found that cluster structure is not disrupted (Fig.~\ref{fig:filter}f).

In conclusion, our analysis demonstrates that semantic embedding can capture relevant characteristics of AI alignment research and that automatic filtering of new publications might be feasible.

\section*{Discussion}
The field of AI alignment research is growing quickly, with many researchers publishing articles on diverse topics. We found that semantic embedding and dimensionality reduction can produce a compact visualization of AI alignment research. This decomposition of AI alignment research mirrors known structures in the field, demonstrating that semantic embedding can capture relevant characteristics of AI alignment research. Furthermore, we demonstrate the possible feasibility of automatically detecting new publications relevant to AI alignment research. In the future, we hope that our decomposition can provide researchers with structured access to the existing literature.

\textbf{Tools for alignment researchers.} Our presented research suggests several exciting possible applications for improving the research landscape in AI alignment research. We have begun to explore this potential by developing several prototypes that use the collected dataset to interactively explore semantic embeddings (Sup.~Fig.~\ref{fig:explorer}), to provide summaries of long articles (Sup.~Fig.~\ref{fig:TLDR}), or to search and compare articles (Sup.~Fig.~\ref{fig:search}). Thanks to the focus on speed and the openness to innovation of the AI alignment research community, we believe that tools tailored to this community might reach broad adoption and help accelerate research efforts.

\textbf{Paradigmatic AI alignment research.} In the language of Thomas Kuhn\autocite{kuhn1970structure}, the successive transition from one paradigm to another via revolution is the usual developmental pattern of mature science. Some researchers argue that AI alignment research is pre-paradigmatic, meaning that it has not yet converged on a single, dominant paradigm or approach. While our research demonstrates that decomposition of AI alignment research into meaningful subfields is possible, we note that the choice of the number of subfields has a subjective component (Fig.~\ref{fig:clustering}d). Furthermore, the semantic similarity between articles in a cluster does not imply similarity in methodology or underlying research agenda. However, we do not believe that this implies the impossibility of progress. In fact, the current exploratory nature of AI alignment research might be a strength, as exploration helps to avoid ossification.

\textbf{Limitations.} Especially due to the rapid expansion of the field (Fig.~\ref{fig:descriptive}), classifications and descriptions of the state-of-the-art might become inaccurate soon after publication. While the observation that our clustering remains stable after including many articles not used for the original clustering (Fig.~\ref{fig:filter}) makes us hopeful, we still plan to carefully monitor the field and publish regular updates to our analysis.

The decision to focus on the two largest, non-redundant sources of articles (Alignment Forum and arXiv) might systematically exclude certain lines of research and thus bias our analysis. However, as a substantial fraction of blog posts, reports and the alignment newsletter tend to be cross-posted or announced on the Alignment Forum we think a strong bias is unlikely.

In summary, by collecting a comprehensive dataset of published AI alignment research literature, we demonstrate rapid growth of the field over the last five years and identify emerging directions of research through unbiased clustering.

\newpage
\section*{Methods}

\textbf{Data collection and inclusion criteria.} 
\begin{itemize}[leftmargin=0.2cm]
    \item \textbf{Alignment Forum \& LessWrong:} We extracted all posts on the forum viewer website \href{https://www.greaterwrong.com/}{\texttt{GreaterWrong.com}} on March 21st, 2022 (dataset used for the analysis in this article) and June 4th (dataset published). We excluded articles with the tag "event", which are published for coordinating meetups.
    \item \textbf{arXiv:} We extended an existing collection of AI alignment research arXiv articles\autocite{AISafetyPapers} from 2020 with relevant publications published since then ("arXiv Level-0"). We started with an existing bibliography of alignment literature\autocite{AISafetyPapers} and augmented that collection with two other bibliographies\autocite{KrakovnaLiterature,LarksLiterature}, articles mentioned in the alignment newsletter, and articles we identified. We excluded articles that were not about AI alignment research.
    \item \textbf{Books:} We converted ebooks into plain text files with \texttt{pandoc}. No text was excluded.
    \item \textbf{Blogs:} We extracted individual articles from AI alignment research-relevant (as determined by the authors) blogs  with the \texttt{requests} and the \texttt{BeautifulSoup} packages. No text was excluded.
    \item \textbf{Newsletter:} We extracted summaries from the publicly available list of summaries and matched them with the respective original articles.
    \item \textbf{Reports:} We extracted additional published articles that were only available as pdf files, by converting these files with \texttt{grobid} and cleaning the resulting files. No text was excluded.
    \item \textbf{Audio transcripts:} We were able to locate some transcripts of interviews available online. For the rest, we used a voice-to-text service (\texttt{otter.ai}) to extract transcripts from AI alignment research-relevant (as determined by the authors) recordings. We hired contractors to clean the resulting transcripts to correct formatting problems and spelling mistakes. After cleaning, no text was excluded.
    \item \textbf{Wikis:} We extracted articles from two open Wikis on AI alignment research (\texttt{arbital.com}, (\texttt{lesswrong.org's} Concepts Portal and \texttt{stampy.ai}) through the export option on the website.
\end{itemize}

\textbf{Data analysis.} We performed the dataset collection with Python 3.7 on commodity hardware and Google Colab and all data analysis with Python 3.7 in Google Colab. We created plots with the seaborn package\autocite{Waskom2021} and post-processed them in Adobe Illustrator. 

\textbf{Semantic embedding.} We used the Allen SPECTER model\autocite{cohan2020specter} through the huggingface sentence transformer library\autocite{reimers-2019-sentence-bert} for embedding articles into a 768 dimensional vector space. The SPECTER model requires each article as \texttt{<Title> + <SEP> + <Abstract>}, where \texttt{<SEP>} is the separator token of the tokenizer. For articles from the arXiv, we used the author-submitted abstract as the \texttt{<Abstract>}. As articles from the Alignment Forum do not always have an author-submitted abstract, we instead used the first 2-5 paragraphs of the article as the \texttt{<Abstract>}.

\textbf{Dimensionality reduction.} To compute a two-dimensional representation of the semantic embedding, we used the python UMAP package\autocite{sainburg2021parametric} with a neighborhood parameter of \texttt{n\_neighborhood=250}. Using a smaller or larger neighborhood did not affect the results, but at very small neighborhood values (\texttt{n\_neighborhood<40}) the embedding became unstable.

\textbf{Unsupervised clustering.} While we explored different clustering algorithms, we eventually converged on the k-means implementation of the scikit-learn package\autocite{scikit-learn}, which is straightforward to interpret while producing robust clustering across multiple instantiations.

\textbf{Statistics.} All statistics were computed with the seaborn package\autocite{Waskom2021} in python, with the exception of the GINI coefficient in Fig.~\ref{fig:interpret}, which we computed as half of the relative mean absolute difference\autocite{enwiki:1088814440}, $${\frac {\displaystyle {\sum _{i=1}^{n}\sum _{j=1}^{n}\left|x_{i}-x_{j}\right|}}{\displaystyle {2n^{2}{\bar {x}}}}},$$
where $x_i$ is the number of articles of each researcher and $\bar x$ is the average number of articles across all researchers.

\textbf{Logistic regression classifier.} To train the AI alignment research classifier, we used the \texttt{LogisticRegression} model of the scikit-learn package in Python\autocite{scikit-learn} with an increased number of maximum iterations, \texttt{max\_iter=1000}. For training, we used $80\%$ of level-0 and level-1 arXiv papers. For evaluation in Fig.~\ref{fig:filter}b we used the remaining $20\%$ of level-0 and level-1 arXiv papers as well as 1000 arbitrarily chosen articles on quantum physics. For the analysis in Fig.~\ref{fig:filter}c-f, we used the arXiv API\autocite{arxivAPI} to collect all articles published in the \texttt{cs.AI} section since its inception.

\textbf{Code and data availability.} The dataset and all code for collecting the dataset is available on Github, \href{https://github.com/moirage/alignment-research-dataset.git}{https://github.com/moirage/alignment-research-dataset.git}. Code for the data analysis is available upon request.

\section*{Acknowledgments}
JK and LR were supported by funding from the Longterm Future Fund.
We thank Daniel Clothiaux for help with writing the code and extracting articles. We thank Remmelt Ellen, Adam Shimi, and Arush Tagade for feedback on the research. We thank Chu Chen, Ömer Faruk Şen, Hey, Nihal Mohan Moodbidri, and Trinity Smith for cleaning the audio transcripts.

\printbibliography
\newpage
\onecolumn
\section*{Appendix} \label{sec:supplementary}
\setcounter{figure}{0}
\renewcommand{\figurename}{Supplementary Figure}

\begin{figure}[H]
    \centerfloat
    \includegraphics[width=0.8\paperwidth]{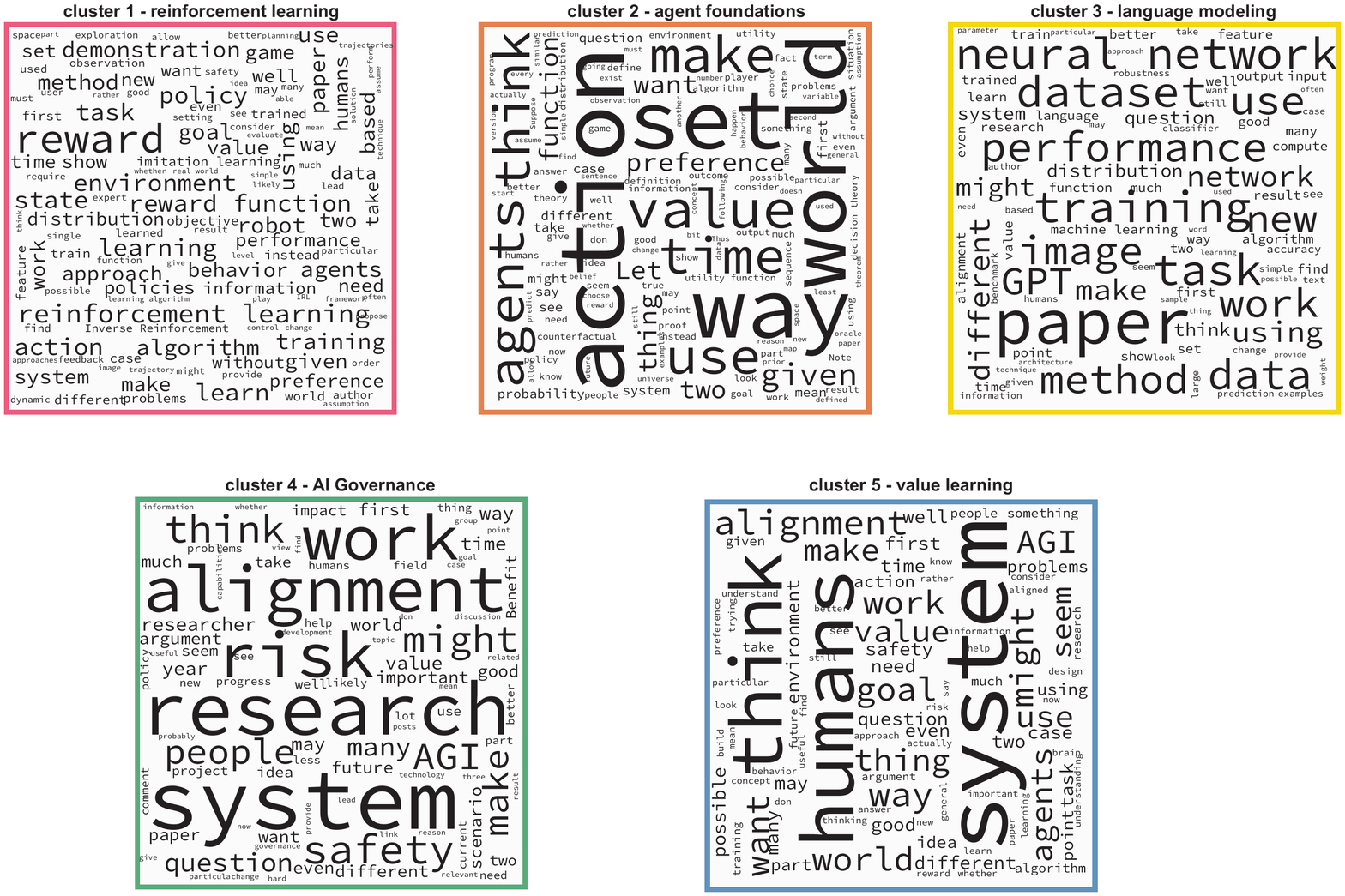}
    \caption[Supplementary Figure]{\textbf{Word frequency visualization for different clusters.} Wordcloud representation (\texttt{word\_cloud} package in Python) of the most commonly used words in articles of the five identified clusters in Fig.~\ref{fig:clustering}. The following words occurred in all clusters very often and were thus removed from the wordcloud: "will", "post", "problem", "example", "one", "SEP", "AI", "agent", "human", "model", and "models". }
    \label{fig:wordcloud}
\end{figure}

\begin{figure}[H]
    \centerfloat
    \includegraphics[width=0.75\textwidth]{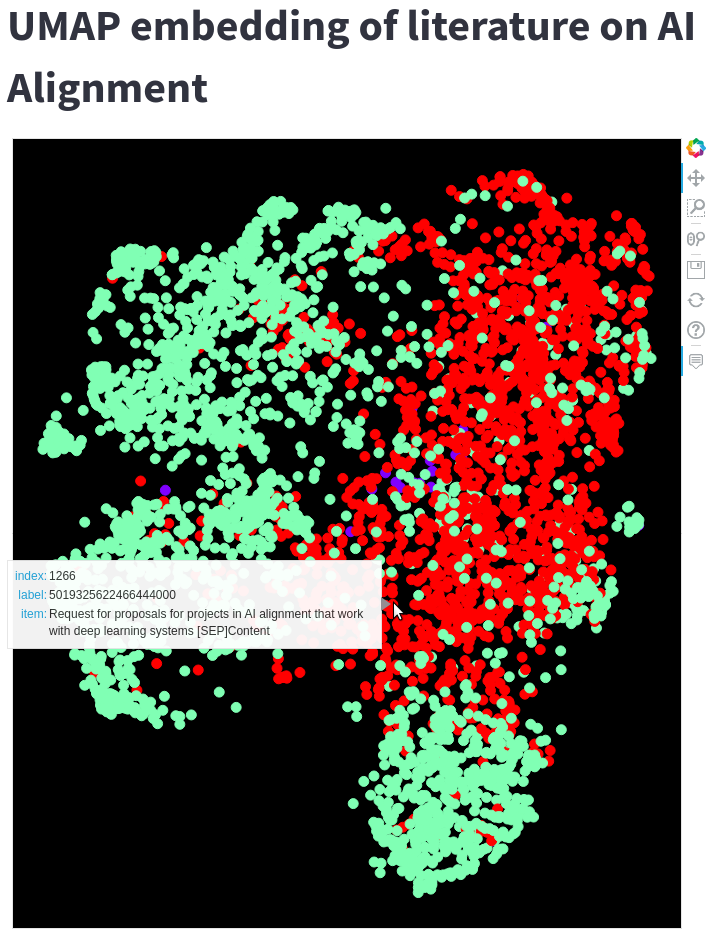}
    \caption[Supplementary Figure]{\textbf{Interactive embedding of AI alignment literature.} An interactive plot (\texttt{plotly.com}) of an UMAP projection of AI alignment research that displays the title of a selected article. }
    \label{fig:explorer}
\end{figure}

\begin{figure}[H]
    \centerfloat
    \includegraphics[width=0.75\textwidth]{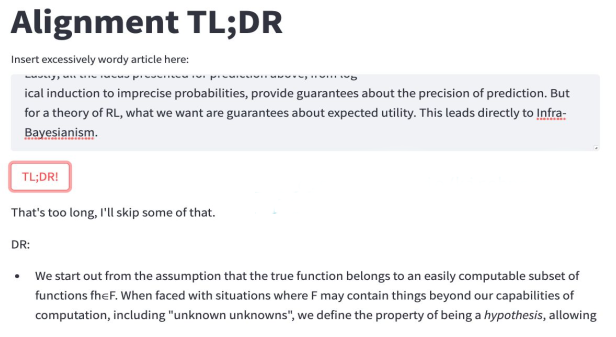}
    \caption[Supplementary Figure]{\textbf{Summarization tool.} An early prototype of a summarization service for AI alignment research articles. We finetuned a 6B \texttt{GPT-J} language model\autocite{mesh-transformer-jax,gpt-j} on the collected dataset and designed a prompt that produces a short summary of a provided AI alignment research article. }
    \label{fig:TLDR}
\end{figure}

\begin{figure}[H]
    \centerfloat
    \includegraphics[width=\textwidth]{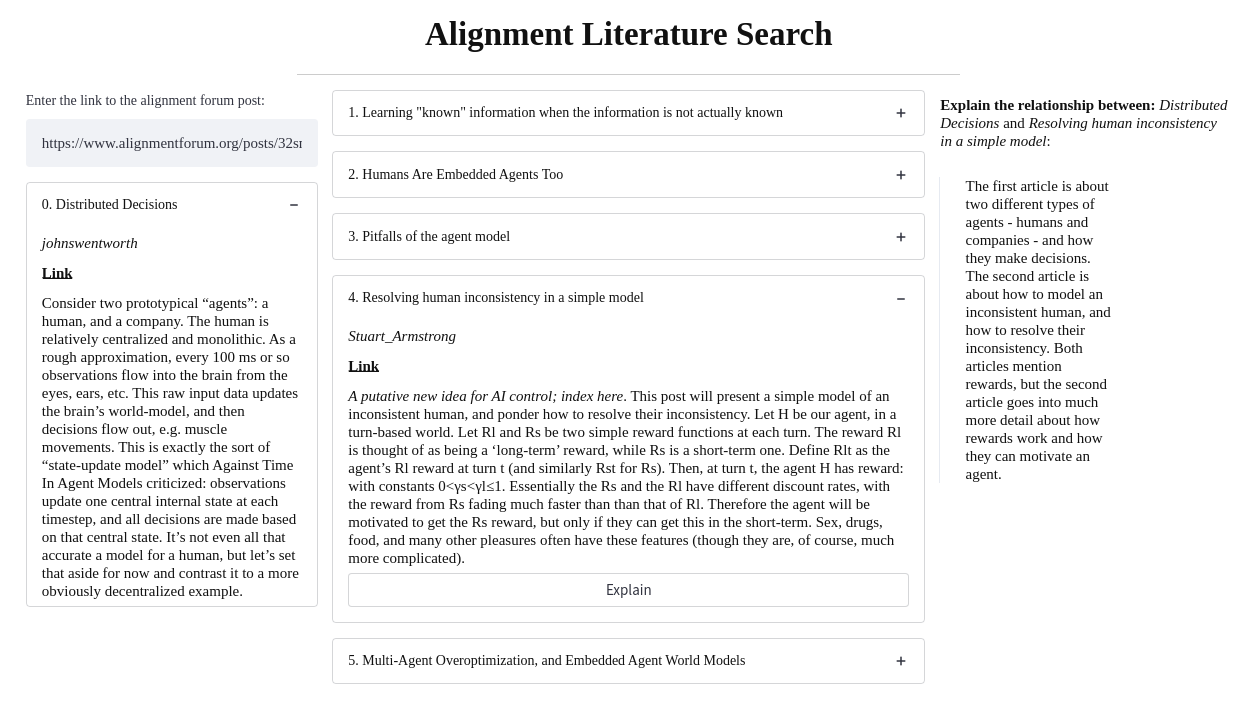}
    \caption[Supplementary Figure]{\textbf{Prototype of semantic search engine.} After entering the URL of an Alignment Forum post (top left), the article is extracted (bottom left) and embedded with the Allen SPECTER model\autocite{cohan2020specter}. The resulting embedding is compared with all embeddings with a vector database search service (\href{https://www.pinecone.io/}{\texttt{Pinecone.io}}) to retrieve similar articles (middle column). By clicking the "Explain" button on a search result, a query with the abstract of the original article and the search result is sent to the OpenAI API to generate an analysis of similarities and differences (right column). }
    \label{fig:search}
\end{figure}

\end{document}